# Xe anions in stable Mg-Xe compounds: the mechanism of missing Xe in earth atmosphere

Mao-sheng Miao*

*Materials Research Laboratory, University of California, Santa Barbara, CA 93106-5050, USA and*

*Beijing Computational Science Research Center, Beijing 10084, P. R. China*

miaoms@mrl.ucsb.edu

The reactivity of noble gas elements is important for both fundamental chemistry and geological science. The discovery of the oxidation of Xe[1-3] extended the doctrinal boundary of chemistry that a complete shell is inert to reaction.[4,5] The oxidations of Xe by various geological substances have been researched[6-10] in order to explain the missing Xe in earth atmosphere.[11-14] Among many proposals,[14-17] the chemistry mechanisms are straightforward as they identify chemical processes that can capture Xe in earth interior. However, all the mechanisms based on current noble gas chemistry face the same difficulty: the earth lower mantle and core are rich in metals and therefore their chemical environment is reductive. On the other hand, up till now, the opposite chemical inclination – the reductive propensity, *i.e.* gaining electrons and forming anions, has not been proposed and examined for noble gas elements.[18] In this work, we demonstrate, using first principles calculations[19] and an efficient structure prediction method,[20,21] that Xe and Kr can form stable Mg-Xe and Mg-Kr compounds under high pressure. These compounds are intermetallic and Xe or Kr is negatively charged. We also find that elevated temperature has large effect in stabilizing Mg-Xe and Mg-Kr compounds. Our results show that the earth has the capability of capturing Xe but not Kr, which is consistent to the depletion of Xe in earth atmosphere. The stability of these compounds suggests that chemical species with completely filled shell may still gain more electrons filling their outer shell in chemical reactions.

Our work is based on accurate density functional calculations, which have been successfully employed in predicting novel compounds and structures under pressure for numerous times.[22-24] Since the structures of these Mg-Xe and Mg-Kr compounds are unknown, we employ a non-biased automatic structure search method based on the particle swarm optimization (PSO) algorithm. The method can search for stable structure across the entire potential energy surface derived from DFT calculations.[20,25] This method has recently been successfully applied to predict structures of many systems.[21,24,26,27]

Six compositions are studied for Mg-Xe and Mg-Kr compounds, including $MgXe_2$, $MgXe$, $Mg_3Xe_2$, $Mg_2Xe$, $Mg_5Xe_2$, $Mg_3Xe$ and similar compounds for Mg-Kr. After a thorough structure search under pressures from 0 GPa to 300 GPa with a step of 50 GPa, we find that the stable Mg-Xe and Mg-Kr compounds always adopt very simple structures, namely the stacked square lattices at the center sites (Figure 1a). The simplest structure of this kind is the CsCl structure, which is adopted by MgXe and MgKr at pressures higher than 100 and 250 GPa, respectively (Figure 1b). It can be viewed as an alternative stacking of Mg and Xe (Kr) square lattices at the center sites. At lower pressure, the double stacking (MgMgXeXe) with space group of P4/nmm is the most stable structure for MgXe and MgKr (Figure 1c). The most stable structure of $Mg_2Xe$ ($Mg_2Kr$) under high pressure consists of a stacking sequence of MgMgXeMgMgXe. The corresponding structure is in I4/mmm space group (Figure 1d). Similarly, the other compositions also consist of the stacking of the Mg and Xe (Kr) square lattices (Supplementary Information Figure S1). Under higher pressure, the Mg and Xe (Kr) layers tends to intercalate into each other, whereas at lower pressure, they tend to segregate, indicating a tendency of decomposition into Mg and Xe (Kr).

Many binary intermetallic compounds are found to adopt a structure of stacked hexagonal lattices. Mg-Xe and Mg-Kr are the first to be found adopting structures of stacked square lattices. Under increasing pressure, not only does the volume of these compounds reduce, but the shape of the unit cell representing by the $c/a$ ratio also changes. Except CsCl structure ($c/a$=1), the cells are longer in $c$ direction under ambient pressure. Increasing pressure reduces the elongation in c direction. For some structures, the unit cell becomes more compressed in $c$ than in $a$ and $b$ directions under very high pressure (Figure S2).

In order to compare the stability of Mg-Xe and Mg-Kr compounds with different composition, the enthalpy of formation per atom is calculated using the following formula for Mg-Xe: $h_f(Mg_nXe_m) = [H(Mg_nXe_m) - nH(Mg) - mH(Xe)]/(n+m)$,[22,26] in which $H$ is the enthalpy of the most stable structure of certain composition at the given pressure. The results are shown in the form of convex hulls in Figure 2, which presents the stability of Mg-Xe and Mg-Kr compounds at high pressures and zero temperature. The compounds located on the convex hull are stable against the decomposition into other compositions, whereas the compounds that are above the convex hull are not stable and will decompose into the compounds located on the hull.

As shown in Figure 2a, Mg-Xe compounds are not stable under pressures lower than 150 GPa. At higher pressure, all Mg-Xe compounds have negative enthalpy of formations. However, only MgXe and $Mg_2Xe$ locate on the convex hull and therefore are stable against the decomposition. Mg-Kr compounds show similar trend, except that the required pressure to stabilize them is much higher, around 250 GPa (Figure 2b). By calculating the phonon spectra, we find that the structures of certain compounds that are stable at the given pressure are also dynamically stable (Figures S4). The effect of the zero-point energies on the enthalpy of formation is found to be negligible, less then 2meV/atom for all the compounds.

Furthermore, we find that temperature can significantly stabilize Mg-Xe and Mg-Kr compounds. The enthalpies of formation of MgXe and MgKr as functions of both pressure and temperature are presented as contour plots in Figure 2c and 2d. At 1000 K, MgXe and MgKr become stable at pressures of 75 and 150 GPa, respectively. Since both temperature and pressure increase with depth of earth, our calculations can be used to examine the stability of these metal-noble gas compounds at different depth. While comparing Figure 2c and 2d with the T, P profiles with earth depth,[28] we find that MgXe can form in the area that is 1500 km below the earth surface, which is in the lower mantle region and is rich in Mg. In contrast, MgKr can only form at about 3000 km below the surface which is close to the outcore and the heavier metal elements such as Fe and Ni start to dominate. Therefore, we conclude that earth has the capability of capturing Xe but not Kr, a noble gas with no report of missing in the earth atmosphere. The reaction of Mg with other noble gas is even harder. Our calculations show that Mg and Ar cannot react below 300 GPa at 0K.

The origin of the stability of Mg-Xe and Mg-Kr can be revealed by their electronic structures. Figure 3a shows the band structure and the projected density of states (PDOS) of MgXe at 100 GPa. The Fermi level crosses several bands in the Brillouin zone, therefore MgXe should be categorized as an intermetallic compound. Interestingly, the states around the Fermi level consist of mainly the Xe 5d states mixed in part with Mg 3s and 3p states. This indicates that the charge transfer from Mg 3s to Xe 5d is the mechanism of forming these compounds under pressure. Using Bader's quantum mechanics atom-in-molecule (QMAIM) charge analysis[29] for MgXe at 100 GPa, we find that the charge transfer is quite large (1.5e per Xe), indicating the compound is strongly ionic. The Bader charge is usually much smaller than the nominal charge, even for typical ionic compounds. For example, the Bader charge of MgO in rocksalt structure at 100 GPa is calculated to be 1.73e, which is only slightly larger than the charge transfer in MgXe. The electron localization function (ELF)[30] is usually used for describing the charge redistribution and the bonding feature of molecules and solid materials. Larger ELF values usually correspond to inner shell or lone pair electrons and covalent bonds, whereas the ionic and metallic bonds correspond to small ELF values. As shown in Figure 3b and 3c, MgXe is quite ionic and there are metallic electrons between the Xe atoms. The chemical nature of MgXe compounds is that there is large charge transfer from Mg to Xe leading to strong ionic interactions between Mg and Xe. The electrons filling Xe 5d orbitals are delocalized in the crystal, making MgXe a 5d metallic compound (intermetallic). Other Mg-Xe compounds and Mg-Kr compounds are also intermetallic, showing similar features as MgXe in their electronic structures.

Xe and other noble gas elements possess a complete atomic shell, therefore are quite inert to chemical reactions. The first Xe compound, $Xe[PtF_6]$, was found by

Bartlett,[1] 30 years after the theoretical prediction of Xe reactivity by Pauling. Since then, many Xe and other noble gas elements compounds are obtained or predicted by reacting them with strong oxidizing agents. In all these compounds, the noble gas elements are positively charged. In a recent paper, we show by DFT calculations that under moderate pressure, Cs can be oxidized beyond +1 state by Fluorine and form $CsF_n$ molecular crystals. The underlying mechanism is that pressure can alter the energy order of Cs 5p and F 2p orbitals and cause the electron transfer and sharing between these states.

The current work shows another effect of pressure, namely reducing a chemical species with completed atomic shells. As shown by our calculations noble gas elements can be reduced and become negatively charged and form stable intermetallic compounds with Mg under pressure. The mechanism of this phenomenon is that the energy of outer shell d orbitals (5d for Xe) increases less significantly under pressure as the 3s orbitals of Mg. If the pressure is high enough, electrons will transfer from Mg 3s to noble gas d orbitals. This effect is more significant for Xe than for Kr and other noble gas elements, which is the reason that earth mantle is capable of capturing Xe but not other noble gases. We actually have seen that Xe can form stable compounds with Li at a pressure much lower than that required for stable Mg-Xe compounds. The effect of the s and d orbital ordering under pressure has been observed in experiments for some alkali and alkaline metals. For example, the s electrons of K will transfer to the d shell under high enough pressure, and K behaves like transition metals with partially filled d bands.[31] Because of the competition of their own outer shell d orbitals, Xe cannot oxidize late alkali and alkaline metals.

**Method**

To obtain stable structures for Mg-Xe and Mg-Kr compounds, we conducted an unbiased structure prediction based on the particle swarm optimisation algorithm as implemented in CALYPSO (crystal structure analysis by particle swarm optimisation).[20,25] The structure predictions were performed using a unit cell containing up to 4 Mg-Xe or Mg-Kr units and at pressures ranging from 0 to 300 GPa. The underlying *ab initio* structural relaxations and the electronic band-structure calculations were performed within the framework of density functional theory (DFT) as implemented by the VASP (Vienna Ab initio Simulation Package) code.[32] The generalised gradient approximation (GGA) within the framework of Perdew-Burke-Ernzerhof (PBE)[33] was used for the exchange-correlation functional, and the projector augmented wave (PAW) potentials[34] were used to describe the ionic potentials. In the PAW potential for Xe, the 5s, 5p, $5d$ and 6s orbitals were included in the valence. The cut-off energy for the expansion of the wave function into plane waves was set at 1200 eV, and Monkhorst-Pack *k*-meshes were chosen to ensure that all enthalpy calculations converged to better than 1 meV/atom.

**References**

1   Bartlett, N. Xenon hexafluoroplatinate (V) Xe⁺PtF₆. *Proceedings of the Chemical Society of London* (1962).

2   Agron, P. A. *et al.* Xenon difluoride and nature of Xenon-Fluorine bond. *Science* **139**, 842 (1963).

3   Braida, B. & Hiberty, P. C. The essential role of charge-shift bonding in hypervalent prototype XeF2. *Nature Chemistry* **5**, 417-422, doi:10.1038/nchem.1619 (2013).

4   Pauling, L. *The nature of the chemical bond.* (Cornell Uni- versity Press, 1960).

5   Murrel, J. N., Kettle, S. F. A. & Tedder, J. M. *The chemical bond.* (John Willey & Sons, 1985).

6     Brock, D. S. & Schrobilgen, G. Synthesis of the Missing Oxide of Xenon, XeO2, and Its Implications for Earth's Missing Xenon. *Journal of the American Chemical Society* **133**, 6265-6269, doi:10.1021/ja110618g (2011).

7     Zhu, Q. *et al.* Stability of xenon oxides at high pressures. *Nature Chemistry* **5**, 61-65, doi:10.1038/nchem.1497 (2013).

8     Sanloup, C. *et al.* Retention of xenon in quartz and Earth's missing xenon. *Science* **310**, 1174-1177, doi:10.1126/science.1119070 (2005).

9     Sanloup, C., Hemley, R. J. & Mao, H. K. Evidence for xenon silicates at high pressure and temperature. *Geophysical Research Letters* **29**, doi:10.1029/2002gl014973 (2002).

10    Sanloup, C., Schmidt, B. C., Gudfinnsson, G., Dewaele, A. & Mezouar, M. Xenon and Argon: A contrasting behavior in olivine at depth. *Geochimica Et Cosmochimica Acta* **75**, 6271-6284, doi:10.1016/j.gca.2011.08.023 (2011).

11    Ozima, M., Podosek, F. A. & Igarashi, G. TERRESTRIAL XENON ISOTOPE CONSTRAINTS ON THE EARLY HISTORY OF THE EARTH. *Nature* **315**, 471-474, doi:10.1038/315471a0 (1985).

12    Sabu, D. D., Hennecke, E. W. & Manuel, O. K. TRAPPED XENON IN METEORITES. *Nature* **251**, 21-24, doi:10.1038/251021a0 (1974).

13    Sabu, D. D. & Manuel, O. K. XENON RECORD OF EARLY SOLAR-SYSTEM. *Nature* **262**, 28-32, doi:10.1038/262028a0 (1976).

14    Jephcoat, A. P. Rare-gas solids in the Earth's deep interior. *Nature* **393**, 355-358, doi:10.1038/30712 (1998).

15    Shcheka, S. S. & Keppler, H. The origin of the terrestrial noble-gas signature. *Nature* **490**, 531-+, doi:10.1038/nature11506 (2012).

16    Bernatowicz, T. J., Podosek, F. A., Honda, M. & Kramer, F. E. THE ATMOSPHERIC INVENTORY OF XENON AND NOBLE-GASES IN SHALES - THE PLASTIC BAG EXPERIMENT. *Journal of Geophysical Research* **89**, 4597-4611, doi:10.1029/JB089iB06p04597 (1984).

17    Bernatowicz, T. J., Kennedy, B. M. & Podosek, F. A. XE IN GLACIAL ICE AND THE ATMOSPHERIC INVENTORY OF NOBLE-GASES. *Geochimica Et Cosmochimica Acta* **49**, 2561-2564, doi:10.1016/0016-7037(85)90124-3 (1985).

18    Grochala, W. Atypical compounds of gases, which have been called 'noble'. *Chemical Society Reviews* **36**, 1632-1655, doi:10.1039/b702109g (2007).

19    Hafner, J. Ab-initio simulations of materials using VASP: Density-functional theory and beyond. *Journal of Computational Chemistry* **29**, 2044-2078, doi:10.1002/jcc.21057 (2008).

20    Wang, Y. C., Lv, J., Zhu, L. & Ma, Y. M. CALYPSO: A method for crystal structure prediction. *Computer Physics Communications* **183**, 2063-2070, doi:10.1016/j.cpc.2012.05.008 (2012).

21    Wang, Y. C. *et al.* High pressure partially ionic phase of water ice. *Nature Communications* **2**, 563, doi:10.1038/ncomms1566 (2011).

22    Feng, J., Hennig, R. G., Ashcroft, N. W. & Hoffmann, R. Emergent reduction of electronic state

dimensionality in dense ordered Li-Be alloys. *Nature* **451**, 445-448, doi:10.1038/nature06442 (2008).

23  Ma, Y. *et al.* Transparent dense sodium. *Nature* **458**, 182-U183, doi:10.1038/nature07786 (2009).

24  Miao, M.-s. Caesium in high oxidation state and as a p-block element. *Nature Chemistry* **accepted** (2013).

25  Wang, Y. C., Lv, J. A., Zhu, L. & Ma, Y. M. Crystal structure prediction via particle-swarm optimization. *Physical Review B* **82**, 094116, doi:10.1103/PhysRevB.82.094116 (2010).

26  Peng, F., Miao, M. S., Wang, H., Li, Q. & Ma, Y. M. Predicted Lithium-Boron Compounds under High Pressure. *Journal of the American Chemical Society* **134**, 18599-18605, doi:10.1021/ja308490a (2012).

27  Zhu, L. *et al.* Spiral chain O-4 form of dense oxygen. *Proceedings of the National Academy of Sciences of the United States of America* **109**, 751-753, doi:10.1073/pnas.1119375109 (2012).

28  Stacey, F. D. & Davis, P. M. *Physics of the Earth*. 4th Edition edn, (Cambridge University Press, 2008).

29  Bader, R. *Atoms in Molecules: A Quantum Theory* (Oxford University Press, 1990).

30  Silvi, B. & Savin, A. Classification of chemical-bonds based on topological analysis of electron localization functions. *Nature* **371**, 683-686, doi:10.1038/371683a0 (1994).

31  Parker, L. J., Atou, T. & Badding, J. V. Transition element-like chemistry for potassium under pressure. *Science* **273**, 95-97, doi:10.1126/science.273.5271.95 (1996).

32  Kresse, G. & Furthmuller, J. Efficient iterative schemes for ab initio total-energy calculations using a plane-wave basis set. *Physical Review B* **54**, 11169-11186 (1996).

33  Perdew, J. P., Burke, K. & Ernzerhof, M. Generalized gradient approximation made simple. *Physical Review Letters* **77**, 3865-3868 (1996).

34  Blochl, P. E. Projector augmented-wave method. *Physical Review B* **50**, 17953-17979 (1994).

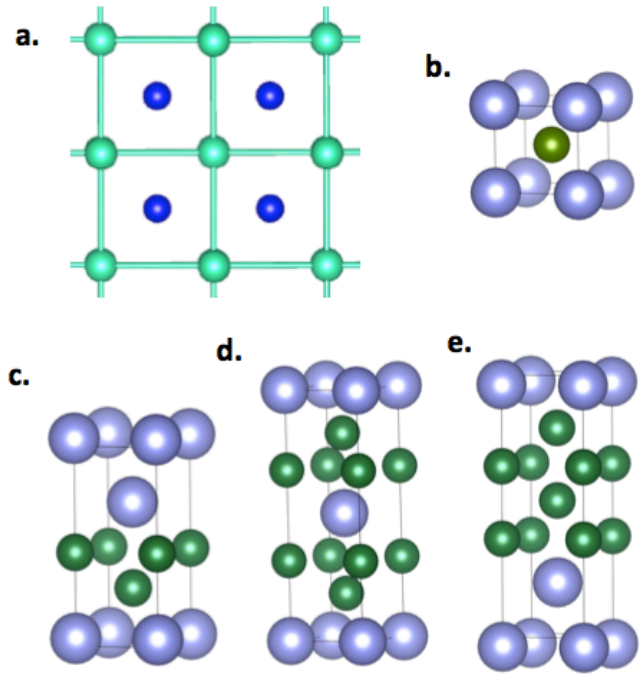

Figure 1. **Selected structures of MgXe and Mg$_2$Xe compounds. a.** Top view of two intercalated square lattices (green-ball and blue-ball lattices). **b.** MgXe in CsCl structure, which is the simplest case of stacked square lattices. **c.** MgXe in *P4/nmm* structure. **d.** Mg$_2$Xe in *I4/mmm* structure. **e.** Mg$_2$Xe in *P4/nmm* structure. In figures **b. – d.**, the large blue balls and the smaller green balls represent the Xe and the Mg atoms.

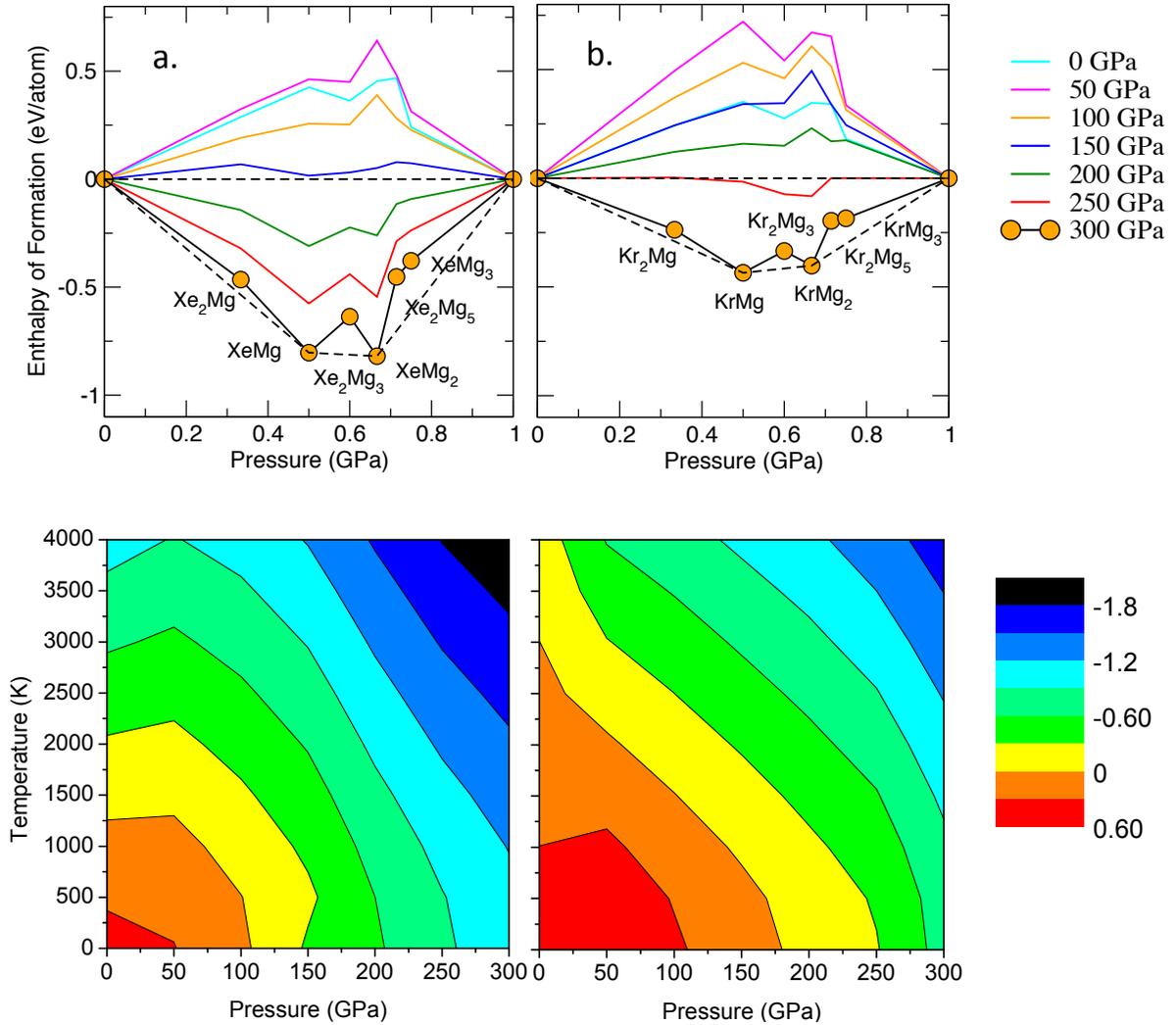

**Figure 2. Stability of Mg-Xe and Mg-Kr compounds under high pressures and elevated temperatures. a.** and **b.** The enthalpies of formations per atom of Mg-Xe and Mg-Kr compounds under a series of pressures. The colors cyan, pink, orange, blue, green, red and black represent the results under pressures of 0, 50, 100, 150, 200, 250 and 300 GPa. The dashed line connecting points for 300 GPa show the convex hull. The compounds locating on the convex hull are stable against decomposition into other compositions. **c.** and **d.** the contour plots of enthalpy of formation of MgXe and MgKr as function of pressure and temperature.

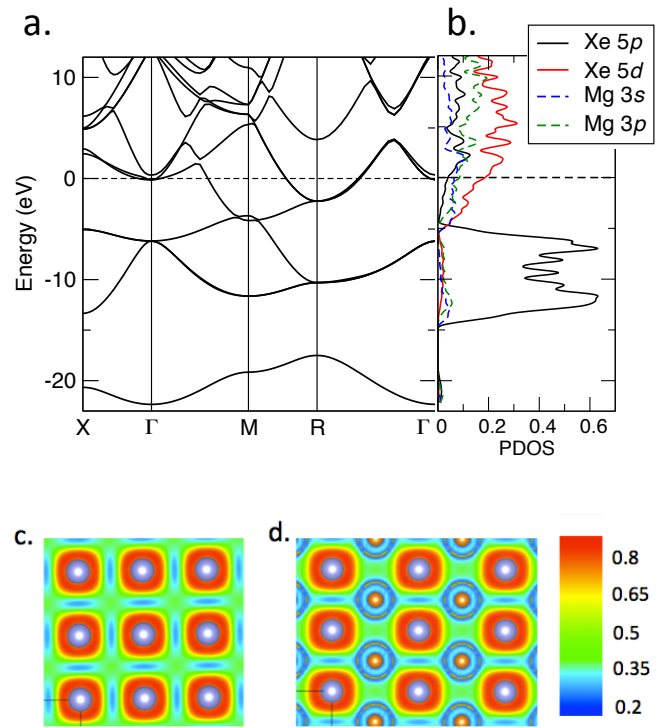

Figure 3. **Electronic structure of MgXe in CsCl structure. a.** Band structure of MgXe under 100 GPa in CsCl structure. **b.** Projected density of states (PDOS) of MgXe under 100 GPa and in CsCl structure. The black and red solid lines represent Xe *5p* and *5d* states; the blue and green dashed lines represent the Mg *3s* and *3p* states. **c.** and **d.** The calculated electron localization functions (ELF) for MgXe in CsCl structure at 100 GPa showing at (100) and (1 -1 0) cutoff planes, respectively.

Supplementary Information

# Xe anions in stable Mg-Xe compounds: the mechanism of missing Xe in earth atmosphere

Mao-sheng Miao

*Materials Research Laboratory, University of California, Santa Barbara, CA 93106-5050, USA and*

*Beijing Computational Science Research Center, Beijing 10084, P. R. China*

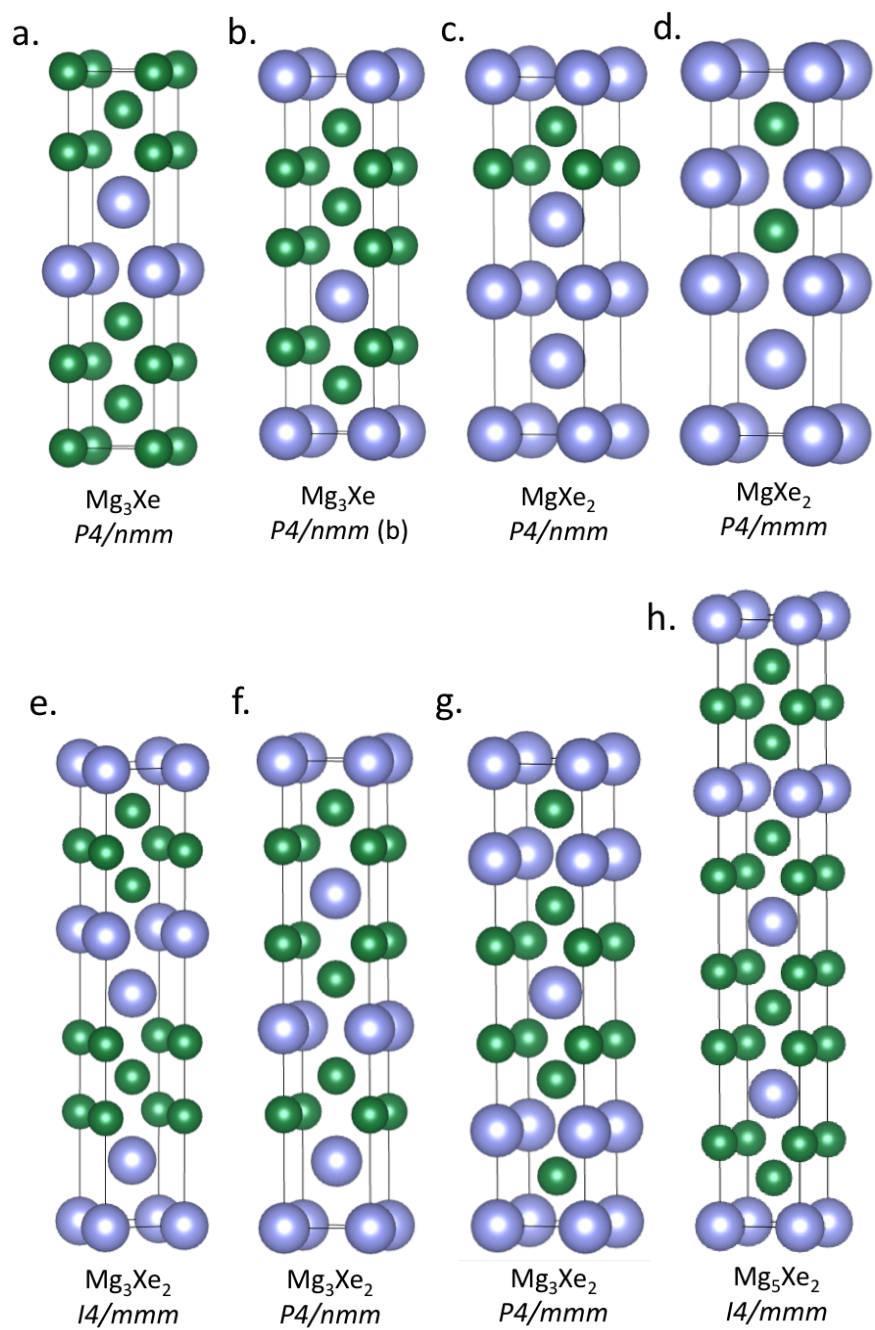

**Supplementary Figure S1. More structures of Mg-Xe compounds. a. and b.** two *P4/nmm* structures for $Mg_3Xe$. **c.** *P4/nmm* structure for $MgXe_2$. **d.** *P4/mmm* structure for $MgXe_2$. **e.** *I4/mmm* structure for $Mg_3Xe_2$. **f.** *P4/nmm* structure for $Mg_3Xe_2$. **g.** *P4/mmm* structure for $Mg_3Xe_2$. **h.** *I4/mmm* structure for $Mg_5Xe_2$. In all figures, the large blue balls and the smaller green balls represent the Xe and Mg atoms respectively.

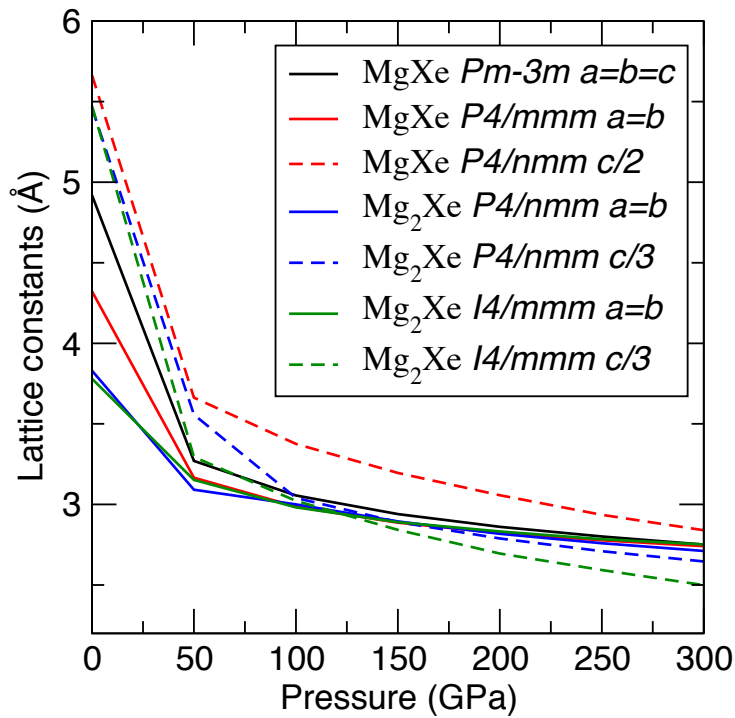

**Supplementary Figure S2. Lattice constants as function of pressure.** $a$ and $b$ are shown as the lattice parameters of the conventional unit cell. $c$ is shown by dividing the number of double layers, *i.e.* the average thickness of every two layers in the conventional unit cell. For CsCl structure, $a=b=c$. For MgXe in *P4/nmm* structure, there are four Mg and Xe layers. $c/2$ shows the average thickness of every two layers. The same is true for $Mg_2Xe$ structures.

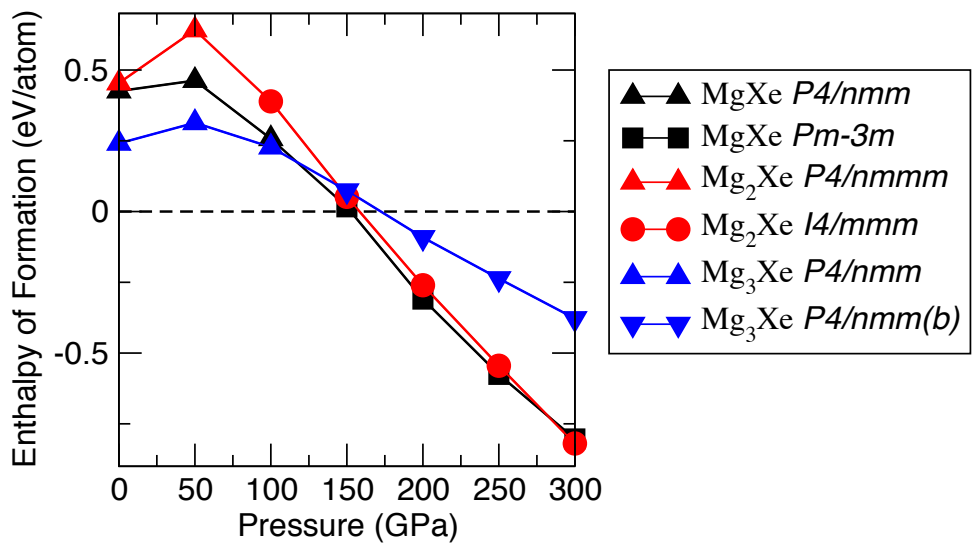

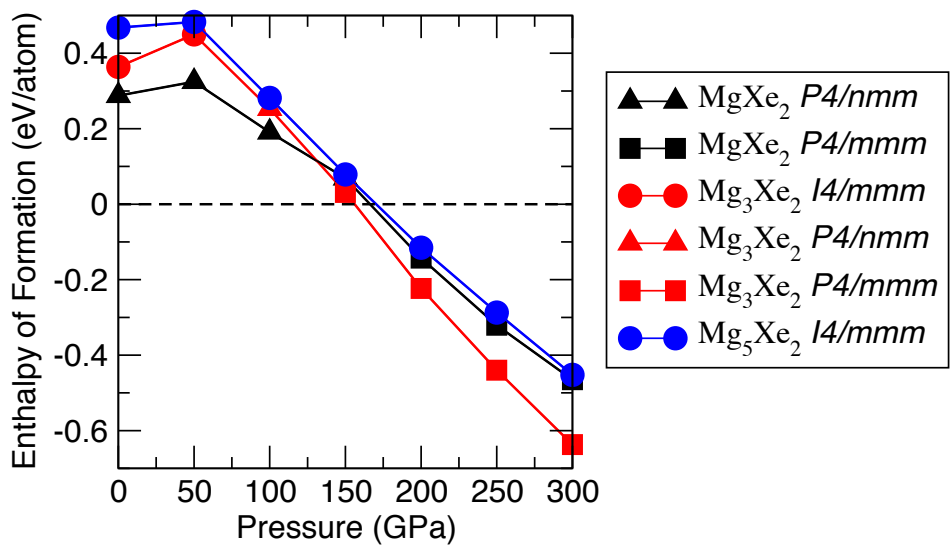

**Supplementary Figure S3. Enthalpy of formation of Mg-Xe compounds as function of pressure.** Upper panel: The enthalpies of formation of MgXe, $Mg_2Xe$ and $Mg_3Xe$. Lower panel: The enthalpies of formation of $MgXe_2$, $Mg_3Xe_2$ and $Mg_5Xe_2$.

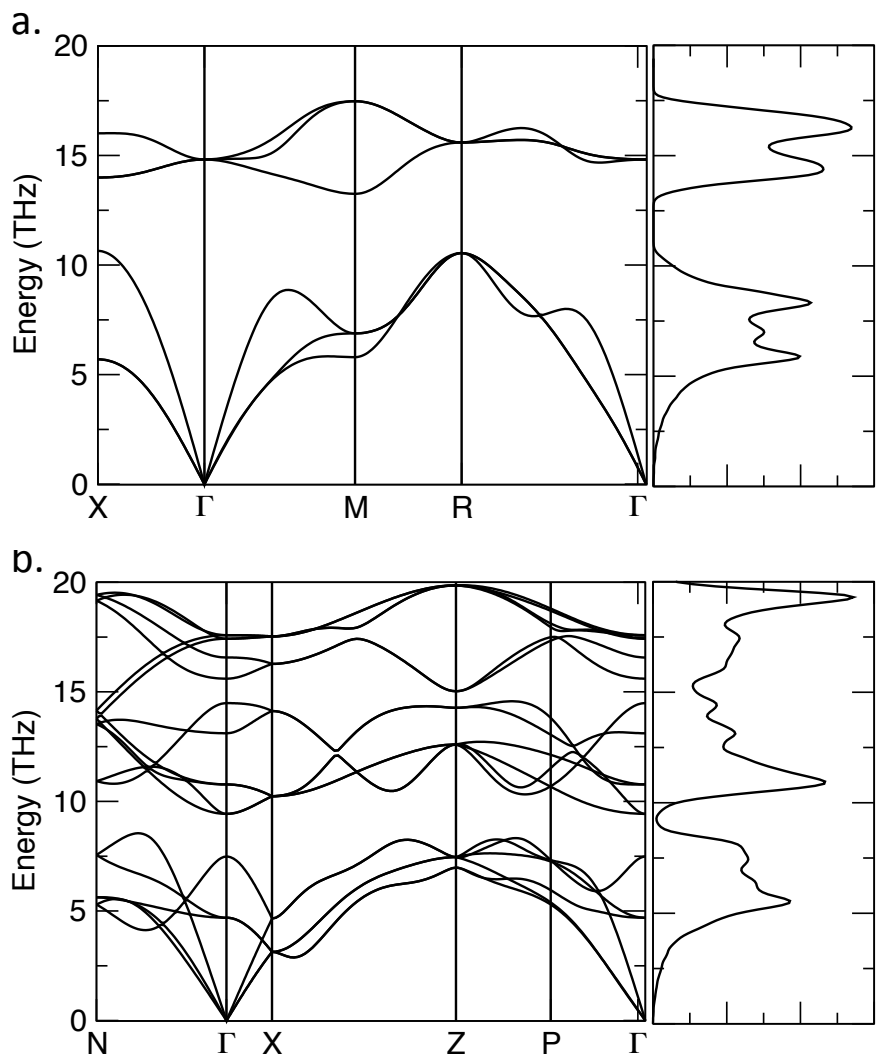

**Supplementary Figure S4. The phonon spectra of MgXe and Mg$_2$Xe at 200 GPa.**
**a.** The phonon dispersion (left) and density of states (right) of MgXe. **b.** The phonon dispersion (left) and density of states (right) of Mg$_2$Xe. The phonon spectra are calculated using finite displacement method by combining the features of phonopy (http://phonopy.sourceforge.net) and VASP programs.